\documentclass{llncs}[12pt]
\usepackage{amsmath}
\usepackage{algorithm}
\usepackage{algpseudocode}
\usepackage{algorithmicx}
\usepackage{graphicx}
\usepackage{float}
\usepackage{hyperref}

\usepackage{tikz}
\usetikzlibrary{automata,positioning, arrows}

\newtheorem{conj}{Conjecture}

\begin{document}
\title{{\em Canonical Number} and {\em NutCracker}: Heuristic Algorithms for  the Graph Isomorphism Problem using Free Energy}
\author{Cewei Cui and Zhe Dang}
\institute{School of Electrical  Engineering and Computer Science\\
Washington State University, Pullman, WA 99164,  USA\\
  \texttt{\{ccui,zdang\}@eecs.wsu.edu}}
\authorrunning{C.\, Cui and Z.\, Dang} 
\maketitle

\begin{abstract}
 This paper develops two heuristic algorithms to solve graph
 isomorphism, using free energy encoding. The first algorithm uses four types of encoding
 refinement techniques such that every graph can be distinguished
 by a canonical number computed by the algorithm. The second algorithm injects energy into
 the graph to conduct  individualization such that the correspondence
 relation between a pair of isomorphic graphs can be found.
 The core principle behind the two algorithms
 is encoding discrete structures as real numbers.
A large  set of experiments  demonstrated  the effectiveness of
 our algorithms.
 \end{abstract}

\pagestyle{plain}

\section{Introduction}
Finding an efficient algorithmic solution to
the graph isomorphism (GI) problem has been  an open problem for more than four decades,
where much research effort has been spent and in the 1970s, there was even a trend called ``Graph Isomorphism Disease" \cite{ReadC77} surrounding the problem.
Why is this problem  so fascinating and attractive to researchers?
The reasons are its obvious importance in
both theory and practice.

In theory, the exact complexity class of GI  is still unknown.
Obviously, it is in NP, but it is unknown whether it is NP-complete.
Many people believe that it is in P, however, no known
  polynomial
time algorithms exist to solve GI in general, though, for some special types of
graphs,   polynomial time algorithms has been shown.
To list a few,  trees \cite{AhoHU74}, interval graphs and planar graphs \cite{ColbournBooth1981},
graphs with bounded genus \cite{Filotti1980}, graphs with bounded degrees \cite{LUKS1982} and
graphs with bounded tree-width \cite{BODLAENDER1990}, do have polynomial time algorithms
for isomorphism.
In practice, graphs are almost the universal data structure
in Computer Science.
In  applications such as bioinformatics and graph mining,
efficient graph matching algorithms are needed. For example, when biologists
and chemists try to query a newly found protein with unknown biological functionalities
 in a
 protein data bank (like RCSB) where each protein in the bank is annotated with its
identified functionalities,  a protein molecule's spatial structure can be used.
An efficient GI algorithm can be used to design an engine
to resolve this query.
In many application areas,  algorithms for the  subgraph isomorphism
are also needed, where a solution to the GI may provide a hint of inspiration to the
widely considered more difficult subgraph isomorphism (which is known to be NP-complete).

In the past 40 years,  there has been some very
 influential research work on the GI problem. To list a few,
 Brendan McKay combined degree-based refinement strategies,
graph automorphisms and search tree to design one of the most useful tools,
{\em Nauty} in \cite{McKay81practicalgraph}
and in \cite{McKayP14} (with Piperno).
L{\'{a}}szl{\'{o}} Babai introduced group theory to solve  GI  in
1970s. His last year's seminal result
has shown that the complexity of the GI problem is quasi-polynomial \cite{Babai15}.
There also exist several important exact match (instead of heuristic)
 graph isomorphism algorithms.
Ullmann algorithm \cite{Ullmann76} uses a backtrack algorithm with some pruning metrics to solve graph
isomorphisms and subgraph isomorphisms. Schmidt and Druffel \cite{Schmidt1976}, in the same year,
invented another backtracking algorithm using distance matrices.
The VF2 algorithm, invented by
Cordella et. al. \cite{Cordella2004}, is also a backtracking algorithm that uses state space representation
to reduce space complexity.

In this paper, we propose efficient and heuristic algorithms to attack the GI problem in a
novel perspective from physics:  thermodynamic formalism.
Thermodynamic formalism provides a mathematical structure to rigorously deduce
a macroscopic characteristic \cite{ruelle2004,sarig1999,walters1982} of a
many-particle system (e.g., gas) from microscopic behavior.
Originally, thermodynamic formalism is mainly used to analyze physical systems and their
mathematical abstractions--dynamic topological systems.  Our recent research \cite{cuidang16}
introduces
 thermodynamic formalism to finite automata, formal languages,
and programs.
In this paper, we use thermodynamics formalism  to
 encode a discrete structure (a graph optionally with weight) into a real number (called potential or energy in physics, and called weight in this paper).
The theoretical underpinning of our approaches is that, when a gas molecule ``walks" on the graph
representing its energy changes along the time, the local potential or weight
assigned on the nodes  or edges will be eventually reflected in its long-term characteristic
(such as equilibrium at infinitely, i.e. the  far future).
The main ideas in our approaches are three-fold:
 (1) local structure such as neighbourhood information can be encoded as a real number;
(2)  the real number generated from neighbourhood information can be assigned as node weights or edge weights;
and (3) global information, such as shortest distance, can also be encoded as edge weights.
Then, we translate a simple graph (i.e., without weights) into a weighted graph.
In this way, spectrum of the weighted graph can be used (heuristically) to tell
 whether graphs are isomorphic.
Stationary distributions can help us to find the correspondence between two isomorphic graphs.

Refinement and individualization is a well known and classic technique to solve graph isomorphism problem
(see \cite{Spielman1996,SpielmanNotes} for an excellent introduction).
In order to make our algorithms easy to understand, avoiding too many tedious mathematical
and physical details, we try to fit our algorithms into the classic refinement and individualization
framework. Hence, our readers can follow the traditional way to understand our algorithms
and also make it easy to compare them to previous algorithms.


\section{The philosophy behind our methods: encoding discrete structures as real numbers}
Although the main mathematical tool used in this paper is thermodynamic formalism,
our path to attack GI problem starts from spectral graph theory, a charming branch of graph theory.
In this paper, we mainly focus on undirected graphs, though the approaches can be straightforwardly generalized to
directed graphs.
A graph $G$ is specified by $(V, E)$ where $V$ is the set of nodes and $E$ is the set of
(undirected) edges (where each edge is an unordered pair of nodes).
Given two graphs $G_1=(V_1,E_1)$ and $G_2=(V_2,E_2)$, they are isomorphic if there is a
one-to-one correspondence mapping $T$
between $V_1$ and $V_2$ such that, for all $u,v\in V_1$,
$(u,v)\in E_1$ iff $(T(u), T(v))\in E_2$.
A basic data structure of a graph is its adjacency matrix $A$.
Spectral graph theory mainly investigates properties of graphs by their spectrum properties,
e.g., their eigenvalues and eigenvectors (of its adjacency matrix).
The theory provides an excellent tool set to the GI problem.  For instance,
 the largest eigenvalue (which is called the spectral radius, or the
 Perron number) remains unchanged when
a one-to-one permutation is applied on its nodes.
Using this theory, many apparently  similar graphs
can be easily distinguished by their  spectra, especially by their largest eigenvalues.
One of the theoretical underpinnings of using the spectral graph theory
for the GI problem is roughly as follows.
Let $\cal X$ be a ``most random" Markov walk on a graph $G$ with its
adjacency matrix  $A$.
The largest eigenvalue of $A$ can tell the entropy rate (which equals
the logarithm of the eigenvalue)
of the  Markov chain while the second largest eigenvalue of $A$ can tell the convergence
rate to the entropy rate
\cite{Cover06}. Both entropy rate and convergence rate
are the long term properties
of a Markov chain. In contrast to this, an edge (or a local structure)
 of the graph resembles one-step of the walk and
hence is a short term property of the Markov chain.

For some nonisomorphic but highly symmetric graphs, it is still difficult to distinguish
their spectra  only
using the adjacency matrices. A more powerful and well-known tool in the spectral graph theory,
Laplacian matrix \cite{SpielmanNotes}, is introduced.  Roughly speaking, Laplacian matrix is another
matrix representation of a simple graph $G$, in the form of $L = D -A$, where $D$ is
degree matrix of $G$ and $A$ is the adjacency matrix of $G$.
The degree matrix $D$ of $G$ is defined as a diagonal matrix, where the $i$-th element
of diagonal is the number of nodes to which $v_i$ is connected.
Analyzing the spectrum of Laplacian matrix $L$ of $G$,   many graphs, which are not distinguishable
under the spectrum of the adjacency matrix  $A$, are now distinguishable.

The improvement made by the Laplacian matrix \cite{SpielmanNotes} over the
adjacency matrix is essentially due to the
additional information brought into the Markov chain's short term behaviors:
the degree information on each node.  In the long term, this information
is eventually reflected in the spectrum of the graph.

However, the Laplacian matrix approach fails on
many  extremely symmetric graphs, such as strongly
regular graphs.  Following the above thread of thinking from
adjacency matrix to Laplacian matrix, we need
 a more ``information rich" matrix representation of  a graph to
further improve the Laplacian matrix approach, by
encoding
  more local structure information to its matrix form.
But, it seems  difficult to achieve this goal at first because
it is not straightforward how to numerically encode a discrete structure
into a real number with a mathematical meaning in the matrix.
Fortunately,
using thermodynamic formalism, we can extract a node's local structure as a subgraph.
Then we compute the free energy of the subgraph, a positive real number, as a weight
on the edge. As a result, the graph's matrix representation is a matrix where each entry
indicates local structure information.  The largest eigenvalue of the weighted matrix
denotes the free energy of the weighted matrix.



In the rest of this section, we briefly introduce the free energy of a graph.  In the next
two sections we present our two heuristic
algorithms for deciding graph isomorphism and finding
correspondence.

Let $G$ be a graph while $\cal X$ is a Markov chain on $G$ that defines a measure $\mu$ on
the set of infinite walks of $G$ (each walk is a sequence of nodes).
The $\mu$ can be defined from the cylinder sets of the walks (we omit the details here; see
\cite{Gurevich1984,sarig-notes}).  We assume that $G$ is connected (i.e., every node
can reach every other node).
Each infinite walk $\alpha=vv'v''\cdots$ carries an energy defined by a function $\psi(\alpha)$.
In a simplest setting, we define $\psi$ on the first step of the walk; i.e.,
$\psi(\alpha)=\psi(v,v')$ and hence the potential function or the energy function
$\psi$ assigns an energy called \textit{weight} to each edge of $G$.
The free energy of $G$, for the given $\psi$, is defined as
$\sup_\mu \{ h_\mu+\int \psi d\mu\},$
where $h_\mu$ is the Kolmogorov-Sinai entropy.
It can be computed using the Gurevich matrix   ${\cal M}$ where
each entry ${\cal M}_{ij}$ of the matrix is $e^{\psi(v_i,v_j)}$ and the natural logarithm of the Perron number of this
nonnegative matrix is exactly the free energy defined earlier. In particular, the unique
$\mu^*$, as well as the stationary distribution, called Parry measure,
$\eta\cdot\xi$ of the Markov chain
defined with the $\mu^*$,
that achieves the supreme can be computed using the left eigenvector
$\eta$
and the right eigenvector $\xi$, after normalization,
 of the Perron number of the matrix \cite{Gurevich1984}.

In the sequel,  for convenience, we directly treat $e^{\psi(v_i,v_j)}$ as the energy or
the weight on an edge  and treat the Perron number, that is the largest eigenvalue
(which is a positive real from the Perron-Frobenius theorem), as the free energy of the graph.

\section{{\em Canonical Number}: a refinement framework using numerical encoding of various discrete structures}
Notably in the previous research on GI, McKay's color labelling refinement approaches and
Schmidt and Druffel's shortest distance matrix are both widely used heuristic metrics to reduce
the size of a search tree for backtracking algorithms. McKay's color-labelling refinement only depends on local structures,
while the shortest distance matrix approach needs a global structure.
Intuitively, if we can encode both local and global structure information into the refinement
procedure, it would be possible to distinguish or assign unique labels to more nodes.
However, this encoding strategy has a fundamental difficulty.
Our  answer uses thermodynamic formalism \cite{ruelle2004}  mentioned earlier, by
numerically encoding a discrete structure as a real number.
Now, we need to exploit which structure information is important for GI problem and how
to numerically encode it on the graph.

\subsection{Shortest distance encoding} \label{sd-encoding}
Shortest distance matrix actually is a well known metric for GI problem.
This metric shows a global constraint on the nodes' relations in a graph.
The theoretical underpinning of this method is the well-known fact that
the shortest distance matrix decides the graph that it represents.
To fit the useful
metric in our context, we make a small modification to it.
For a connected graph, the shortest distance matrix is straightforward.
For a graph that is not connected, when there is no path between two given nodes,
the distance is defined as infinity (which is set to be a
large number that
only needs to be  larger than the diameter of the graph).
Then,
 we use the reciprocal
 of the shortest distance between nodes $i$ and $j$ as the weight on the edge
between $i$ and $j$. At first,
the ``reciprocal"-representation of
the shortest distances may seem counter-intuitive.
The reason we modify it this way is that after doing so,
 every graph becomes connected, which is necessary in
thermodynamics formalism (i.e., in this way, the Gurevich matrix is
now irreducible for any graph).


\subsection{Neighbourhood graph encoding} \label{nh-encoding}
It is known that neighbourhood is an important local structure that can distinguish
many similar graphs. For example, in the class of strongly regular graphs (16,6,2,2),
there exist two non-isomorphic graphs. The core difference between the two non-isomorphic
graphs is that the neighbourhood in one graph is of two triangles while the neighbourhood
in the other graph is a $6$-cycle graph. Thus, if we can encode this type of structural
differences,
many types of graphs can be easily distinguished.

First, we need to define what is a neighbourhood graph.
Given a graph $G=(V, E)$, for a node $v_i$ in the node set $V$, we say $G_{i}^{\tt nh}=(V_i^{\tt nh}, E_i^{\tt nh})$ is
the neighbourhood graph of $v_i$ if every node in $V_i^{\tt nh}$ is a node connected
(with an edge)  to $v_i$ in $G$
and any two nodes $v_x$ and $v_y$ in $V_i^{\tt nh}$ in $G_{i}^{\tt nh}$ is connected iff
$v_x$ and $v_y$ are connected in $G$.

To encode the neighbourhood graph as a real number we need to overcome two challenges:
\begin{itemize}
\item The subgraph is not necessarily connected. How can we compute the energy
of an disconnected graph?
\item After computing the energy of the neighbourhood graph,
where can we assign the energy?
\end{itemize}

To solve the first challenge, we compute its standard shortest distance matrix first.
Following the modification for shortest distance matrix in the previous subsection,
now, we have an irreducible matrix and then we compute the Perron number as the
free energy of the neighbourhood graph. For the second challenge,
after we compute the free energy of the neighbourhood graph, we add
 the energy to the existing weight (which now is the reciprocal  shortest distance) of  every
edge between $v_i$ and other nodes (including $v_i$ itself).


\subsection{Shared neighbours subgraph encoding} \label{sn-encoding}
In strongly regular graphs, the third parameter such as in (16,6,2,2)
 tells us the number of shared neighbours between
two adjacent nodes while the fourth parameter indicates  the number of shared neighbours between
two adjacent graph.  Hence,
the shared neighbour is a useful metric to demonstrate the local structure
of different nodes pairs.

Now, we need to define the shared neighbours graph.
Given a graph $G=(V, E)$, for a node $v_i$  and $v_j$ in the node set $V$,
we say $G_{ij}^{\tt sn}=(V_{ij}^{\tt sn}, E_{ij}^{\tt sn})$ is
the shared neighbours graph of $v_i$ and $v_j$ if every node in $V_{ij}^{\tt sn}$ is a node connected (with an edge) to both
$v_i$ and $v_j$ in $G$
and any two nodes $v_x$ and $v_y$ in $V_{ij}^{\tt sn}$ in $G_{ij}^{\tt sn}$ is connected iff
$v_x$ and $v_y$ are connected in $G$.

Given a pair node of $v_i$ and $v_j$, the encoding procedure is in the following:
\begin{itemize}
\item Obtain the shared neighbours graph for the node pair $v_i$ and $v_j$, $G_{ij}^{\tt sn}$.
\item Compute the free energy of $G_{ij}^{\tt sn}$ and add the free energy to the weight
 on the edge between $v_i$ and $v_j$ on $G$.
\end{itemize}

\subsection{Union neighbours subgraph encoding} \label{un-encoding}
Shared neighbours graph is a powerful metric to inspect graph's characteristics.
But, the metric has some fundamental limitations. For example, given a bipartite graph $G$, all nodes of $G$
can be divided into two blocks $V_{left}$ and $V_{right}$. Following the definition of bipartite graph,
obviously, any node in $V_{left}$ shares zero neighbour with any node in $V_{right}$. Hence,
shared neighbours graph encoding is not
enough.
Our solution is union neighbours, which is also quite natural
from the view of Venn diagrams.
 We assume that the neighbourhood of node $i$ is $A$ and
the neighbourhood of node $j$ is $B$, the shared neighbour can be understood as the intersection of
$A$ and $B$. Hence, if we use  intersection operation on sets, naturally, it follows naturally
 that
we also need a union operation on sets, i.e., the union of $A$ and $B$, to complete a Venn diagram.

Now, we need to define union neighbours graph.
Given a graph $G=(V, E)$, for a node $v_i$  and $v_j$ in the node set $V$,
we say $G_{ij}^{un}=(V_{ij}^{un}, E_{ij}^{un})$ is
the union
 neighbours graph of $v_i$ and $v_j$ if every node in $V_{ij}^{un}$ is a node connected to either
$v_i$ or  $v_j$ in $G$
and any two nodes $v_x$ and $v_y$ in $V_{ij}^{un}$ in $G_{ij}^{un}$ is connected iff
$v_x$ and $v_y$ are connected in $G$.

Given a pair node of $v_i$ and $v_j$, the encoding procedure is in the following:
\begin{itemize}
\item Obtain the union neighbours graph for the node pair $v_i$ and $v_j$, $G_{ij}^{un}$.
\item Compute the free energy of $G_{ij}^{\tt un}$ and  add the free energy to
 the weight
 on the edge between $v_i$ and $v_j$ on $G$.
\end{itemize}

\begin{algorithm}
\begin{algorithmic}[1]

\footnotesize

\Require{$A$, which is the adjacency matrix of $G$ and the dimension of $A$ is $n$.}
\Statex
\Function{Perron}{$M$}
\State Compute the Perron number (i.e. the largest eigenvalue) of a matrix $M$, $\lambda$.
\State return $\lambda$
\EndFunction

\Statex

\Function{CanonicalNumber}{$A$}
\State $n \leftarrow$ the dimension of $A$.
\State Build an $n \times n$ matrix $W=\{W_{ij}\}$ with initial values zero.
\State Compute the shortest distance matrix \{$shortest\_distance_{ij}$\}.

 \For{$i$, $j$ in \{$W_{ij}$\}}
    \State $W_{ij} \leftarrow \frac{1}{shortest\_distance_{ij}}$
 \EndFor
\State Construct  $1 \times n$  matrix NHTable with zero as initial values.
\State Construct two $n \times n$ matrices SNTable and UNTable with zero as initial values.
 \For{$i$, $j$ in \{$W_{ij}$\}}
    \State Obtain the shared neighbours graph of node pair $(i,j)$: $G_{ij}^{\tt sn}$
    \State Compute the Perron number of the graph $G_{ij}^{\tt sn}$:  $SNTable_{ij} \leftarrow Perron(G_{ij}^{\tt sn})$
    \State Obtain the union neighbours graph of node pair $(i,j)$: $G_{ij}^{un}$
    \State Compute the Perron number of the graph $G_{ij}^{un}$: $UNTable_{ij} \leftarrow Perron(G_{ij}^{un})$
    \If{$i==j$}
        \State Obtain the neighbourhood graph of node $i$: $G_{i}^{\tt nh}$
        \State Compute the Perron number of graph $G_{i}^{\tt nh}$: $NHTable_{i} \leftarrow Perron(G_{i}^{un})$
    \EndIf
 \EndFor
 \For{$i$, $j$ in \{$W_{ij}$ \}}
    \State $W_{ij} \leftarrow W_{ij}+ NHTable_{i} + NHTable_{j} + SNTable_{ij} + UNTable_{ij} $
 \EndFor
 \State Use the weight matrix $W$ as the matrix representation of $G$
 \State Compute the Perron number of graph $G$: $canonical\_number = Perron(W)$
 \State return $canonical\_number$.
  \EndFunction

  \Statex
\State CanonicalNumber($A$).
  \end{algorithmic}

 \caption{{\em Canonical Number}: Graph Isomorphism Testing Algorithm}
 \label{CN-algorithm}
 \end{algorithm}

\subsection{The Algorithm {\em Canonical Number}}
The pseudocode of our algorithm is listed in Algorithm \ref{CN-algorithm}.
Inspired by the canonical labelling method of
previous GI research, we name our algorithm
the  {\em Canonical Number} algorithm.
For a given graph $G$, there exists a canonical labeled graph $G_C$ such that any graph $G'$ that
is isomorphic to $G$ shares the same canonical labeled graph $G_C$.
Our experiments presented in the latter section show that
for all the graphs that we run our algorithm upon,
 any two nonisomorphic graphs
have  different numbers and any isomorphic graphs share the same number.
This characteristics is similar to canonical labelling, hence we call the number, the
free energy of the graph returned from the algorithm,
the canonical number of the graph.

\subsubsection{Correctness}
To prove the correctness of our algorithm, we need to show two parts:
\begin{itemize}
\item For any isomorphic graphs, they share the same canonical number.
\item For any two nonisomorphic graphs, they have different canonical numbers.
\end{itemize}
The first part is obvious because the encoding process does not depend on
the naming of any node.
For now, we cannot mathematically prove the second part.
So we call our algorithms a
heuristic algorithms. However, we have the following conjecture:

\begin{conj}
There exists at lease one encoding method that assign weights on graphs such that
all isomorphic graphs share the same canonical number and any nonisomorphic graphs
have different canonical numbers.
\end{conj}
We also conjecture that our current encoding method is one of these encoding methods
(and hence we conjecture  that GI is in P).

\subsubsection{Explanation of the algorithm}
We present a brief explanation to  Algorithm \ref{CN-algorithm}.
In lines 1-4, the function Perron is used the compute Perron number, i.e., the largest eigenvalue,
of a given matrix.
In line 5-30, the function CANONICALNUMBER is designed to compute the canonical number of a given
graph $G$.
In lines 6-11, we implement the shortest distance encoding described in Subsection \ref{sd-encoding}.
In lines 12-13, we initialize three tables to store values for neighbourhood, shared neighbours
and union neighbours encoding.
In lines 14-23, we iterate every edge in the graph to conduct various encoding approaches.
In lines 15-16, shared neighbours encoding, described in Subsection \ref{sn-encoding}, is implemented for the edge between node $v_i$ and $v_j$.
In lines 17-18, union neighbours encoding, described in Subsection \ref{un-encoding}, is implemented for the edge between node $v_i$ and $v_j$.
In lines 19-22, neighbourhood encoding, described in Subsection \ref{nh-encoding}, is implemented for node $v_i$.
In lines 24-26, we use previous encoding tables to update the matrix $E$, such that $E$ is a matrix
representation of $G$.
In lines 27-28, we compute the free energy of $E$ and return it as the canonical number of $G.$

\subsubsection{The time complexity analysis of our algorithm}
Algorithm \ref{CN-algorithm} includes several basic algorithms.
The time complexity of computing shortest distance \cite{floyd1962algorithm} is $O(n^3)$;
The time complexity of computing eigenvalues of a matrix with dimension $n$ is
as same as the matrix multiplication. For the ease of discussion, we use $O(n^3)$ as
its complexity although it is may a litter lower than this.
In Algorithm \ref{CN-algorithm}, the dominating part of the time complexity is the loop
in lines 13-22. In the for-loop, the complexity of one iteration depends on obtaining
a subgraph and computing the subgraph's Perron number, i.e., $O(n^2+n^3)=O(n^3)$.
The complexity of the for loop itself is $O(n^2)$. Hence, the time complexity of
this algorithm is $O(n^5)$.

\section{{\em NutCracker}: a free energy based refinement-and-individualization for finding correspondence}
In the previous section, we proposed an heuristic approach to do refinement.
Then the canonical number of the refined matrix was used to identify different graphs.
Can we just use the refinement procedure to find the correspondence relation
between a pair of isomorphic graphs? To achieve this goal, the
refinement procedure needs to be powerful enough such that every node has a unique label.
For some non-symmetric graphs, this is possible. However, for many very symmetric graphs,
it is almost impossible. For example, given a pentagon graph $G$, if we conduct
the previous refinement methods, we still cannot assign unique labels for every node.

\begin{figure}[H]
\caption{Nuts and singles in the  NutCracker algorithm}
\centering
\includegraphics[scale=0.3]{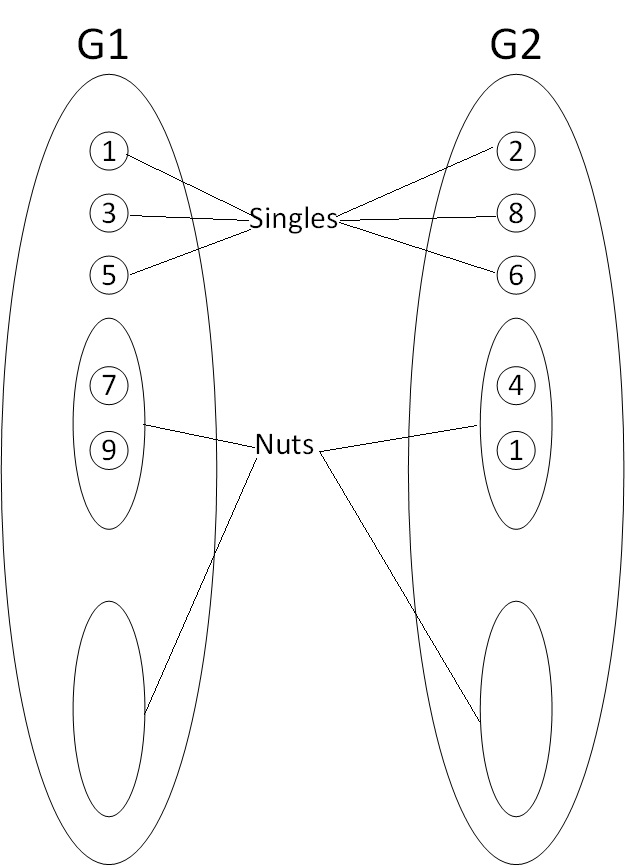}
\label{nutcracker-example}
\end{figure}

It is true for many graphs  that many nodes cannot have a unique label after refinement.
In past GI research, there are  two well-known methods (see \cite{SpielmanNotes}
for a clear introduction) to solve this issue.
First,  one builds a search tree to search every possible case with some additional heuristics.
Second, one tries to find a distinguishing set such that if all nodes in the distinguishing set
are labeled first, then all the other nodes can get their unique labels through refinement.
The second method is called {\em individualization}.
Both of the methods  may lead to exponential complexity.

Individualization is absolutely necessary to break the symmetry of the graph.
However, do individualization methods always require a distinguishing set?
Our approach and experiments show that
this may not be required. In order to obtain a polynomial time individualization method,
we try to design a step-by-step individualization approach where in  every step only
one node is individualized. This approach implicitly assumes (this assumption is actually
true for all the experiments we have done) at every step,
we can always choose the correct node and never need to backtrack.
As a result, assuming $G$ has $n$ nodes,
we need, at most, $n$ steps to finish the individualization procedure.

Now, we present the general idea of our algorithm as follows.
First, we need to  use the refinement procedure in the previous section
and obtain a weight matrix $W$.
Second, we need to compute aforementioned
Parry measure of $W$.
Notice that from \cite{walters1982}, given an irreducible
$W$, the Parry measure is unique.
Then, we use its stationary distribution (on nodes) to distinguish different nodes.
See Figure \ref{nutcracker-example}.
After computing the Parry measure, we call a node {\em single} if the node has unique
stationary probability and call a set of nodes {\em nut} if every node in the set shares
the exactly same stationary probability.
In Fig. \ref{nutcracker-example}, node $1,3$ and $5$ in $G_1$ and node $2,8$ and $6$
are singles; nodes $7$ and $9$ in $G_1$ form a nut while node $4$ and $1$ in $G_2$ form
a nut.

For every single in $G_1$ and $G_2$, it is easy to use its stationary probability
to build the correspondence. The obstacle is how to build the correspondence between
nodes in the nuts. Since $G_1$ and $G_2$ are isomorphic,
every correspondent nut should share exactly the same size
and stationary probability.
Then, it is easy to find a pair of such nuts, say $nut_1^{G_1}$ and $nut_1^{G_2}$, where $nut_1^{G_1}$ is in $G_1$
and $nut_1^{G_2}$ is in $G_2$.
In $nut_1^{G_1}$, we randomly choose a node $v_1$ while a node $v_2$ is also randomly chosen
from $nut_1^{G_2}$ such that if we assign a unique integer number to both
$v_1$ and $v_2$ as node weight, the canonical numbers of $G_1$ and $G_2$
are still same.
(In fact,  $v_1$ and $v_2$ are not randomly chosen, see the following pseudo-code step 11
of the Algorithm \ref{nutc}.)
 Hence, we assign the unique integer number to $v_1$ and $v_2$.
The node weight of a node $v$ can be assigned to each edge between
node $v$ and all other nodes. Why did we do it in this way?
Here we use a node as a cracker in a nut to crack the whole graph.
The assigned node weight can be understood as an energy cracker.
The energy cracker injects
a small amount of energy into this node.
Then the amount of energy we just injected will spread to
all other nodes in the entire graph.
After a moment, the dynamic system represented by the graph will
enter a new equilibrium state.  At this step, we check every node's new stationary distribution.
Obviously, any previous single node is still single, and at least one  node in nuts
may become single. Then, we check if any nuts still exist.
If so, we repeat the energy injection procedure; otherwise, every node has becomes single,
i.e. every nodes has a unique label. Obtaining all unique labeled nodes, it is easy to use
the unique labels to build the correspondence relation between nodes in two graphs.


\begin{algorithm}
\footnotesize

\begin{algorithmic}[1]
\Require{$G_1=(V_1, E_1)$ and $G_2=(V_2, E_2)$ are a pair of isomorphic graphs}
\Statex
\State Initialize an array $nodeweight1$(resp. $nodeweight2$) for $G_1$(resp. $G_2$) with zero as initial values.
\State Using $nodweight1$ (resp. $nodeweight2$), assign node weights to every node in $G_1$ (resp. $G_2$).
\State For every node $v$ in $G_1$ (resp. $G_2$), we spread $v$'s node weights to edges between $v$ and
all nodes (including $v$). (Notice that, after this step, $G_1$ and $G_2$ are weighted graphs.)
\State Using the weighted matrix representation $W_1$(resp. $W_2$) of $G_1$ (resp. $G_2$),
compute Parry measure for $G_1$(resp. $G_2$).
\State Through Parry measure, we can easily obtain the stationary distribution $P_1$(resp. $P_2$) of $G_1$(resp. $G_2$).
\State Perform partition on $V_1$(resp. $V_2$) by its values in $P_1$(resp. $P_2$).
\begin{itemize}
\item If a node in $G_1$ (resp. $G_2$) has a unique stationary probability,
we add it to $singlesList1$(resp. $singleList2$).
Then, $singlesList1$(resp. $singleList2$)is obtained.
\item If several nodes in $G_1$ (resp. $G_2$) share the same probability, we combine these nodes to form a nut and
add the nut to the $nutList1$(resp. $nutList2$). Hence, $nutList1$(resp. $nutsList2$) is obtained.
\end{itemize}
\State If all nodes are singles, we are successful; otherwise, continue.
\State Sort the $singleList1$(resp. $singleList2$) by stationary probability.
 Sort $nutList1$ (resp. $nutList2$) by the size of nut. (After this step, $singleList1$ and $singleList2$ should
 be aligned, i.e., the entry in $singleList1$ and $singleList2$ in the same position should have
 same stationary probability.)
\State Create an empty array $newNodeweigh1$(resp. $newNodeweight2$) for $G_1$(resp. $G_2$).
\State From the starting position to the end position of both singleLists ($singleList1$ and $singleList2$),
we pick the nodes in the same position from two lists, position by position. We assign
a unique weight to both nodes. Herein, The weight is unique to the position, rather than
two lists. ($newNodeweigh1$ and $newNodeweight2$ are updated).

\State For the first nut in $G_1$, $nut_1^{G_1}$ and the first nut in $G_2$, $nut_1^{G_2}$,
we pick a node $v_1$ in $nut_1^{G_1}$ and a node $v_2$ in $nut_1^{G_2}$ such that, if we assign
an unique weight to $v_1$ and $v_2$, the updated weighted graph $G_1$ and $G_2$ share the exactly
same canonical number.
\State If we cannot find $v_1$ and $v_2$, the algorithm fails.
\State We update the $nodeweight1$  and $nodeweight2$: $nodeweight1 \leftarrow newNodeweigh1$,
$nodeweigh2 \leftarrow newNodeweigh2$
\State Go to Line 2, repeat the procedure.

\Statex
\end{algorithmic}

\caption{{\em NutCracker} algorithm}\label{nutc}
\end{algorithm}

\subsection{Correctness and time complexity}
\subsubsection{Correctness}
Similar to Algorithm \ref{CN-algorithm}, we cannot mathematically prove the correctness of
this {\em NutCracker} algorithm (otherwise, GI is polynomial).
Hence, we call this algorithm an heuristic algorithm.  The correspondence found by the
algorithm can be easily verified whether it is indeed an isomorphic correspondence.
For the graphs that we run our experiments, the algorithm is indeed correct.

\subsubsection{Time complexity}
For the {\em NutCracker} in Algorithm \ref{nutc}, the denominating part is in Line 11.
From the section, it is known that CanonicalNumber algorithm is $O(n^5)$.
In line 11, the canonical number algorithm is called in a nested loop.
The worse case, we may need to loop $O(n^2)$ times. So, the line 11 is $O(n^7)$.
Also, Algorithm \ref{nutc} is a recursive algorithm, in the worst,
case, we may call it $n$ times. Thus, the worst time complexity is $O(n^8)$.
Note that, this algorithm seems to be a backtrack algorithm.  But, we assume at every step,
we can find the correct choice; we never backtrack. If we make
the wrong choice, the algorithm fails, rather than backtrack.

\section{Experiments}
In this subsection, we present a large set of experiments to demonstrate the
effectiveness of our heuristic algorithms in practice.

\subsection{Experimental subjects}
In order to demonstrate the effectiveness of our algorithms, it is ideal that we carry out
experiments on all available public datasets. However, it is impossible for us
to do so due to limited resources (time and available high performance computers).
So, we choose to run our algorithms on
both ``easy" graphs where other heuristic  algorithms may also succeed
and ``challenging" graphs where other heuristic algorithms  may fail.
 The number of graphs that we run on must also be large enough to have a practical meaning.
Finally, we decide to choose the following datasets for the experiments:
\begin{itemize}
\item The dataset of all connected graphs with exactly 10 vertices
that is maintained by McKay\cite{10nodesgraphs} and generated by his famous program Nauty.
The dataset contains roughly 11,000,000 non-isomorphic  graphs;
\item Strongly regular graphs are known to be notoriously difficult for all GI algorithms.
Spence \cite{srgdata}
 maintains
a dataset that includes all strongly regular graph with the number of vertices less than
or equal to 64. The dataset contains roughly 40,000 non-isomorphic  graphs.
\end{itemize}

\subsection{Experimental Setup}
Our experiments are designed to validate the following two questions.
\begin{itemize}
\item
\textbf{Question 1.}
In our graph isomorphism testing algorithm,  we claimed that
every graph obtains a unique canonical number generated by our algorithm.
Is our algorithm effective for all graphs in the datasets?
\item
\textbf{Question 2.}
Can our NutCracker algorithm find the correspondence between two isomorphic graphs?
\end{itemize}

We design our first experiment to answer Question 1 as follows.
First, we run our canonical number  algorithm on all datasets and generate
an entry for each graph in the format of index number and the canonical number.
Second, for all graphs with the same vertices, we sort them by the
canonical numbers and find the most
similar pairs where the similarity is the difference between its canonical numbers.
For a pair of graphs, the less the distance is, the more similar they are.

We design our second experiment to answer Question 2.
First, for every graph in our datasets, we generate a random permutation
and apply the permutation to the original graph to generate an isomorphic graph.
A pair of isomorphic graphs are then  obtained.
Second, we run our {\em NutCracker} algorithm on every isomorphic graph pair. If we
find the correspondence, we verify that the correspondence is indeed an isomorphic
correspondence between the two isomorphic graphs in the pair.
However, if the correspondence found by the algorithm  fails the verification,
an error is output.

Our two programs are written in Java, with roughly 1,500 lines of code each.
We carry out all our experiments on a single node
of Washington State University's high performance computing cluster.
The node is an IBM dx360 computer which consists of
six cores with 2.67GHz and 24 GB physical memory.

\subsection{Results}
\subsubsection{{\em Canonical Number} algorithm}
For strongly regular graphs, every graph in this dataset obtained  a unique
canonical number from our algorithm. The program implementing
the algorithm took $\approx 12$ hours to compute the entire dataset of
the 40,000 strongly regular graphs.

For the dataset of  11,000,000   connected graphs of ten nodes, every graph also obtained a
unique canonical number from our algorithm.
The program  took $\approx 26$ hours to compute the entire dataset.
But, canonical numbers
of a small number of
 graphs had an extremely small difference.   For these graphs, it
 is better to use a multi-precision version of our program
 to re-verify them (N.b. Java does not have a multi-precision matrix library).
We re-implemented the algorithm in MATLAB and used its high-precision symbolic
package to verify that, for these graphs, the canonical numbers were indeed canonical
(hence our algorithm is correct for the dataset).

\subsubsection{{\em NutCracker} algorithm}
For strongly regular graphs, every graph in this dataset can find the isomorphic correspondence
successfully, using the Java program implementing our algorithm.
 It took $\approx 12$ hours to compute the entire dataset of 40,000 strongly regular graphs.
For the dataset of
11,000,000 connected graphs of ten nodes, under the Java double number
precision using 8 digits of scientific representation, the program can
find the isomorphic correspondence successfully
for every pair of the graph in the dataset and its randomly permuted
version, except for two graphs in the dataset.
 Applying  9 digits of precision, the algorithm finds the correspondence of the two graphs
successfully.
The algorithm took $\approx 26$ hours to compute through the entire dataset of
11,000,000 graphs.

\subsection{Discussions}
In the above experiments, our algorithms are correct for all the datasets. But,
 it is still possible that there
exists some graphs where our heuristic algorithms
may fail. The readers are welcome to use our algorithms to
identify a counter example to help us more deeply understand the GI problem.

In our algorithms we use the free energy, which is a real number,
 to distinguish between all non-isomorphic
graphs. It is known that most programming languages, such as Java, only
provide double float numbers to conduct computation on real numbers.
The eigen-decomposition uses complicated numerical algorithms.
In the process, the precision error may be propagated from one stage to another.
We spent some efforts to minimize such precision errors.
But, fundamentally,
Java needs a higher-precision
matrix library to avoid such errors.

\section{Applications and Future work}
\subsection{Graph mining}
As the popularity of social networks and fast development of knowledge graphs grows,
graph mining will become an emerging research focus of data mining. In graph mining,
there are at least two important
research topics. One problem is mining frequent or common subgraphs and
the other is graph clustering.
For mining frequent or common subgraphs, it is possible that there exists a type of
weight assignment encoding such that, after applying the encoding technique, the number of
frequent subgraphs contained in a graph can be indicated by the free energy.

For graph clustering, one of the most widely used algorithm is the Markov clustering algorithm, which
depends on flow simulation \cite{Dongenthesis}. Applying our canonical number algorithms to graph clustering problem,
the clustering problem become the simplest clustering problem--one-dimensional clustering.
Also, we need to
point out that the weight assignment methods are
not unique and can be adapted  to real
world requirements.
\subsection{Cheminformatics and Bioinformatics}
In cheminformatics, it is an important topic that, given the spatial structure of a
compound, it is efficient to find the compound in a database that only stores spatial structures.
In bioinformatics, we only need to replace the compound  with a protein.
Our algorithms in this paper  can be used to implement a tool to resolve such queries.

\subsection{Future work}
Our future work consists of  several parts.
First, we need to mathematically prove the correctness of our algorithms or find the
limitation (i.e. a counter example) of our algorithms.
Second, we will try to find simpler and more efficient free energy
encoding strategies.
Third, we will apply our theory to solve real world problems in graph mining,
cheminformatics, and bioinformatics.

\section*{Acknowledgements}
We would like to thank  William J. Hutton III for  discussions.
We also would like to thank WSU High Performance Computing for providing resource to carry out
our experiments.

\bibliography{gi}{}
\bibliographystyle{plain}

\end{document}